\documentstyle[12pt]{article}

\newcommand{\be}{\begin{equation}}
\newcommand{\ee}{\end{equation}}
\newcommand{\bea}{\begin{eqnarray}}
\newcommand{\eea}{\end{eqnarray}}

\begin{document}
\begin{titlepage}

\flushright{IP-BBSR-2010-02 }

\vspace{1in}

\begin{center}
\Large
{\bf DUALITY SYMMETRY OF STRING THEORY: A WORLDSHEET PERSPECTIVE}

\vspace{1in}

\normalsize

\large{  Jnanadeva  Maharana\\
E-mail: maharana$@$iopb.res.in \\
( October 2, 2010 )
  }

\normalsize
\vspace{.5in}

 {\em Institute of Physics \\
Bhubaneswar - 751005 \\
India  \\ }

\end{center}

\vspace{1in}

\baselineskip=24pt
\begin{abstract}
We study duality and local symmetries of closed bosonic string
from the perspectives of worldsheet approach in the phase space path integral
formalism. It is shown that the Ward identities reflecting the local
symmetries associated with massless excitations such as graviton and 
antisymmetric tensor can be cast in a duality covariant form. It is shown
how the manifestly $O(d,d)$ invariant Hamiltonian can be obtained in the
Hassan-Sen toroidal compactification scheme, $d$ being the number of compact
dimensions. It is proposed that massive excited states possess a T-duality
symmetry for constant (tensor) backgrounds. This conjecture is verified
for the first massive level.

\end{abstract}

\vspace{.7in}

\end{titlepage}

%\section{Introduction }

%\setcounter{equation}{0}

%\def\theequation{\thesection.\arabic{equation}}
One of the marvels of the string theory is its rich symmetry contents
and notable among these are the dualities. The underlying string dynamics in
diverse dimensions is primarily understood through the web of dualities which
unravel intimate connections between different string theories. It is
recognized that target space duality, the T-duality, can be tested in
perturbation theory. When we consider evolution of a string in the background
of its massless excitation in the first quantized approach, the worldsheet
action is expressed as a 2-dimensional 
$\sigma$-model action and the massless backgrounds
play the role of coupling constants. The vanishing of the corresponding
$\beta$-functions lead to the "equations of motion" of the those backgrounds.
The string effective actions have played very important role in
understanding of string theory from several perspectives. Moreover, if we adopt
 toroidal compactification and require that the backgrounds do not depend on
these compact coordinates, then the reduced effective action manifests the
associated T-duality symmetry. The target space duality is also understood
from the worldsheet point of view. In this approach, we associate a dual
coordinate for every compact direction of the string coordinate and derive
equations of motion for each of the set. Furthermore, with suitable
combination of the two sets,  equations of motion can be expressed in a 
manifestly duality covariant form. \\
The string effective action is known to be invariant under target space local
symmetries such as general coordinate transformation, associated with the
graviton, vector gauge transformation, associated with the two-form
antisymmetric field and nonabelian gauge transformations in the
presence of  nonabelian
massless gauge fields which appear  in certain
compactified theories. There is a proposal to unravel these local
symmetries from  the worldsheet view point.\\
The purpose of this letter is two fold. It is argued that, at least for closed
bosonic string, the Hamiltonian description manifestly exhibits the duality
symmetries. Furthermore, we derive Ward identities intimately related to
the symmetries of the afore mentioned massless states of the string which
are covariant under duality transformations. We adopt the phase space 
Hamiltonian formalism to derive these results. These will be stated more
precisely in sequel. 
Furthermore, we present some evidence that excited massive string states also
exhibit duality symmetry. These are similar to $R\leftrightarrow {1\over R}$
duality symmetry. At this stage, we can verify our conjecture when the higher
dimensional backgrounds (tensors) are constant.\\
Let us consider a closed bosonic string in the background
of its massless excitations graviton, $G_{MN}$,  and antisymmetric tensor 
$B_{MN}$,  where target spacetime indices, $M,N=1, 2, ...{\cal D}$.
\bea
\label{action}
S={1\over 2}\int d\sigma d\tau \bigg( \gamma ^{ab}{\sqrt {-\gamma}}
G_{MN}(X)\partial_a X^M\partial_b X^N+\epsilon ^{ab}B_{MN}(X)\partial _a X^M
\partial _b X^N \bigg)
\eea
Here $X^M(\sigma ,\tau)$ are string coordinates and 
$\gamma ^{ab}$ is the worldsheet metric. 
The classical action is invariant
under worldsheet coordinate reparametrization. A simple example of worldsheet
duality symmetry is to consider flat target space metric and set $B_{MN}=0$.
The spectrum is invariant under $\sigma \leftrightarrow \tau$ which amounts
to $P_M \leftrightarrow X'^{M}$, $P_M$ being the canonical
momenta, prime and 'overdot', denote derivatives with respect to $\sigma$ and $\tau$
respectively. Moreover, if one compactifies a spatial
coordinate of a closed string on $S^1$ with radius $R$, the perturbative
spectrum matches with that of another string if the corresponding 
 coordinate is compactified on a circle of radius
$1\over R$ when we interchange the Kaluza-Klein modes with the winding modes 
and $R \leftrightarrow {1\over R}$; subsequently this symmetry has been studied 
in more general
settings \cite{revodd, prc}.  When some of the spatial coordinates of a  string are 
compactified on torus, $T^d$, d being the number of compact directions with constant 
backgrounds
$G_{\alpha\beta}$ and $B_{\alpha\beta},~\alpha , \beta =1, 2, ..d$, the duality
 group is $O(d,d, {\bf Z})$, ${\bf Z}$ being integers.  
If the backgrounds assume only time dependence, the string effective action is
expressed in a manifestly $O( D, D)$ invariant form, where $ D$ 
is the number of
spatial dimensions \cite{mv} which has interesting consequences in string 
cosmology \cite{gmv,prc}. In a more generalized setting  one adopt a
toroidal compactification scheme when the target space manifold $\cal S$
is decomposed to ${\cal S}=S_{spacetime}\otimes K$ where $D=0, 1,.. D-1$ are 
the spatial
dimensions and $K=T^d$ with $D+d={\cal D}$.
Furthermore, if the backgrounds $g_{\mu\nu},~b_{\mu\nu},~ \mu ,\nu=0,1,..D-1$ 
and $G_{\alpha\beta},~B_{\alpha\beta},~ \alpha ,\beta =1,2,..d$ depend only on 
the spacetime coordinates $x^{\mu}$, 
then the reduced effective action is expressed in a manifestly $O(d,d)$ 
invariant form \cite{ms}. It is worth while to recall some of the salient
features of  T-duality from the worldsheet perspective. It was shown by Duff
\cite{duff}, for constant backgrounds $G$ and $B$, that the evolution equations
of the string coordinates can be cast in an $O({\cal D}, {\cal D})$ covariant
form. For each string coordinate $X^{M}$, he introduced a dual set of 
coordinates ${\tilde Y}^{M}$ and expressed the equations of motion of the
$2\cal D$ coordinates in the duality covariant form. For the next simplest
case, if $G$ and $B$ assume time dependence it was shown that the worldsheet
equation of motion can be expressed in an $O({\bf D},{\bf D})$ covariant
form where ${\bf D}$ are the number of spatial dimensions \cite{j}. 
On this occasion,
for each spatial string coordinate, $X^I$, I being the  spatial index, 
a dual coordinate ${\tilde Y}^I$ was introduced and combined 
equations were cast  in  manifestly 'duality' covariant form.  \\
The worldsheet approach to T-duality for toroidal  compactification was 
addressed by Schwarz and JM \cite{ms} in a general frame work and it was
demonstrated that by, introducing dual coordinates along compact dimensions,
an $O(d,d)$ covariant worldsheet equations of motion can be derived. 
Subsequently, Siegel has advanced these ideas in another direction, introducing
the two vierbein formalism and extending them to supersymmetric theories
\cite{siegel}.\\
Let us briefly recapitulate essentials of phase space Hamiltonian formalism
and reformulate the problem in a duality invariant frame work.
The two constraints associated with $\tau$ and $\sigma$ reparametrization 
respectively are
\bea
\label{constraint1}
{\cal H}_c&=&{1\over 2}\bigg(P_MP_NG^{MN}+X'^{M}X'^{N}G_{MN}-
P_MG^{MP}B_{PN}X'^{N}\nonumber\\
&+& X'^{M}B_{MP}G^{PN}P_N -B_{MP}G^{PQ}B_{QN}X'^MX'^N\bigg)\simeq 0\nonumber\\
 && P_MX'^M \simeq 0
\eea
${\cal H}_c$ is the canonical Hamiltonian derived from (\ref{action})
These are primary constraints which vanish weakly, derived without any
specific choice of the worldsheet metric, $\gamma^{ab}$. 
In order to express them
in a duality invariant form, let us combine $P_M$ and $X'^{M}$ to define a 
$\cal D$-dimensional $O({\cal D}, {\cal D})$ vector
\bea
\label{oddvector}
{\cal V} =\pmatrix{P_M \cr X'^{M}\cr}
\eea
The canonical Hamiltonian density can be re-expressed as
\be
\label{hamiltonian}
{\cal H}_c= {1\over 2}{\cal V}^T{\cal M}{\cal V}
\ee
in a matrix notation where ${\cal M}$ is a $2{\cal D} \times 2{\cal D}$ matrix
\cite{mmatrix}
\bea
\label{matrix}
{\cal M} = \pmatrix {G^{-1} & -G^{-1} B \cr
BG^{-1} & G - BG^{-1} B\cr} \eea
where $G$ and $B$ stand for backgrounds $G_{MN} (X)$ and  $B_{MN}(X)$
appearing in (\ref{action}). 
Note that the equal $\tau$ canonical Poisson Bracket (PB) relation 
\bea
\label{pb}
\{X^{M}(\sigma), P_N(\sigma ') \}_{PB}=\delta ^M _N\delta 
(\sigma - \sigma ')
\eea
translates to 
\bea
\label{zpb}
\{{\cal V}(\sigma), {\cal V}(\sigma ') \}_{PB}=\eta{d\over{d\sigma '}}
\delta (\sigma -\sigma ')
\eea
where 
$\eta =\pmatrix{0 & {\bf 1} \cr {\bf 1} & 0 \cr}$, the 
${2\cal D} \times 2{\cal D}$ matrix is the $O({\cal D}, {\cal D})$ metric,
 ${\bf 1}$ being the ${\cal D}\times{\cal D}$ unit matrix. 
Under the global $O({\cal D},{\cal D})$ 
transformations
\bea
\label{omega}
{\cal M}\rightarrow \Omega{\cal M}\Omega^T, ~~\Omega ^T \eta\Omega=\eta, ~~
\Omega\in O({\cal D}, {\cal D})
\eea
The linear combinations of the above two constraints (\ref{constraint1}) 

\bea
\label{lplus}
{\cal L}_{\pm}= {1\over 2}{\cal H}_c \pm {1\over 4}{\cal V}^T\eta {\cal V} 
\eea
satisfy the equal $\tau$ PB algebra
\bea
\label{vira}
\{{\cal L}(\sigma)_{\pm},{\cal L}(\sigma ')_{\pm} \}_{PB} \simeq \pm
\bigg({\cal L}(\sigma)_{\pm}+{\cal L}(\sigma ')_{\pm}\bigg)
{d\over{d\sigma}}\delta (\sigma -\sigma ') \eea
and 
\label{zeropb}
\bea
\{{\cal L}_+(\sigma), {\cal L}_-(\sigma ') \}_{PB} = 0
\eea
Therefore, ${\cal L}_\pm (\sigma)$ are a pair of first class constraints and
the theory is covariantly quantized adopting Fradkin-Vilkovisky Hamiltonian
formalism \cite{fv}. In the context of closed string, in the background of its
massless excitations, the Hamiltonian phase space BRST quantization was carried
out by us \cite{v,mv}. The corresponding BRST charge is obtained by adopting
the standard procedure
\be
\label{brs}
{\cal Q}_{BRST}=\int d\sigma \bigg[{\cal L}_+\eta _++{\cal L}_-\eta _-+
{\cal P}_+\eta _+\eta '_+ -{\cal P}_-\eta _-\eta '_- \bigg]
\ee
Here the pair of ghosts $\{ \eta _+,\eta _- \}$ are introduced, as is the
prescription, 
for the two first class constraints $\{ {\cal L}_+ , {\cal L}_- \}$ which 
depend on ${\cal V}$ and the backgrounds, ${\cal M}$. 
$\{ {\cal P}_+, {\cal P}_- \}$ are conjugate 
ghost momenta. The gauge fixed Hamiltonian density 
${\cal H}_{\zeta}=\{\zeta , {\cal Q}_{BRST} \}_{PB}$. 
For choice of orthonormal gauge:
$\zeta= {\cal P}_+ +{\cal P}_-$ and 
\bea
\label{gauge}
{\cal H}_{ON}={\cal L}_++ {\cal L}_-+2{\cal P}_+\eta '_++{\cal P}'_+\eta _+
-2{\cal P}_-\eta '_- -{\cal P}'_-\eta _-
\eea
This was the starting point to derive Ward Identities (WI) associated with the 
massless states of the closed string. As alluded to above, we intend to obtain
similar WI in a duality covariant manner. The first step is to introduce the
Hamiltonian action 
\be
\label{haction}
S_H=\int d\sigma d\tau \bigg[P_M{\dot X}^M - {\cal H}_{ON}\bigg]
\ee
In order to unravel the symmetry encoded due to general coordinate 
transformation invariance which is  intimately related
to the presence of graviton, let us consider a generating functional 
\be
\label{gravity}
{\cal Q}_G=\int d\sigma P_{M}\xi^M(X(\sigma)) 
\ee
responsible for an infinitesimal transformation,  $\xi^M(X)$ being the 
parameter.
The variations of phase space variables, ghosts and the $O({\cal D}, {\cal D})$
 vectors are obtained by evaluating their PB with ${\cal Q}_G$ i.e.
\bea 
\delta_{{ \cal Q}_G} { \cal V} =\{ {\cal V}, {\cal Q}_G \}_{PB}, ~~   
\delta_{{ \cal Q}_G}{ \eta _{\pm}}=0, ~~ \delta_{{ \cal Q}_G}{\cal P}_{\pm}=0 
\eea
and in particular $\delta X^M =\xi ^M(X)$; indeed ${\cal Q}_G$ induces general
coordinate transformations. Since the arguments of $G_{MN}$ and $B_{MN}$ are
shifted their variations under (\ref{gravity}) are
\bea
\label{gvary}
\delta _{{\cal Q}_G}G_{MN}(X)&=& G_{MN},_P(X)\xi^P(X), ~~~~
\delta _{{\cal Q}_G}G^{MN}(X)=
 G^{MN},_P(X)\xi^P(X),\nonumber \\ 
&& \delta _{{\cal Q}_G}B_{MN}(X)= B_{MN},_P(X)\xi^P(X)
\eea 
comma stands for  the ordinary derivative here and everywhere.
Thus the components, ${\cal M}^{MN}, {\cal M}^M_N ~ {\rm and}~ M_{MN}$, 
of the ${\cal M}$-matrix
being  functions of $X$ also transform according to the above prescriptions.
The variation of the action is
\bea
\label{qtrans}
\delta _{{\cal Q}_G}S_H\sim \int d\sigma \bigg[{1\over 2}\delta _{{\cal Q}_G} 
{\cal V}^T
{\cal M}{\cal V} +{1\over 2} {\cal V}^T \delta _{{\cal Q}_G} {\cal M}{\cal V}
+{1\over 2} {\cal V}^T{\cal M}\delta _{{\cal Q}_G}{\cal V}
\bigg]
\eea 
It is a straight forward and tedious calculation to check that
\bea
\label{sgct}
\delta_{{\cal Q}_G} S_H=-\delta _{GCT}S_H
\eea
The {\it r.h.s.} of the above equation is to be interpreted as follows. The
Hamiltonian action, $S_H$ depends on ${\cal M}$, expressed in terms of
$G_{MN}, G^{MN}, ~ {\rm and }~ B_{MN} $.  These tensors transform according to
the rules given below \cite{gctr}
\bea
\label{gct}
\delta_{GCT}G_{MN}&=&-G_{MP}\xi^P,_{N}-G_{PN}\xi^P,_{M}-G_{MN},_P\xi^P
\nonumber\\
 \delta_{GCT}B_{MN}&=&-B_{MP}\xi^P,_N-B_{PN}\xi^P,_M-B_{MN},_P\xi^P
\eea
The next step is to define the Fradkin-Tseytlin generating functional, 
$\Sigma$, in the phase space Hamiltonian path integral formalism \cite{ft}
\bea
\label{ftz}
\Sigma [{\cal M}]=\int {\cal D}P{\cal D}X{\cal D}\eta _{\pm}{\cal D} {\cal P}_
{\pm} e^{iS_H[P, X', \eta _{\pm}, {\cal P}_{\pm}, {\cal M} ] }
\eea
Notice that under canonical transformations, the phase space measure is 
invariant, at least classically. The issue of noninvariance of this measure,
which might lead to anomalies, will be touched upon briefly later. Moreover, if
we implement the canonical transformation (\ref{gravity}) on 
$\Sigma[{\cal M}]$ and the variation of $S_H$ under (\ref{gravity}) is
compensated through (\ref{sgct}), as was argued in \cite{v,mv}. Then
 we arrive at 
\bea
\label{wi1}
\delta_{GCT}\Sigma[{\cal M}]=\bigg<\int d^Dx{{\delta S_H}
\over{{\delta\cal M}(x)}}
\delta{\cal M}(x) \bigg>_{{\cal M}}=0
\eea
In the above equation $<...>$ is to be understood as the functional integral
weighed with $exp(-iS_H)$. Note that the functional derivative of the action,
$S_H$, with respect to the background is the corresponding vertex operator.
Therefore, (\ref{wi1}) translates to 
\bea
\label{wi2}
\bigg<\int d^2\sigma \delta (x-X(\sigma))V_{\cal M}^{PN}
\bigg({\cal M}_{PR}\xi^{R},_N+{\cal M}_{RN}\xi^{R},_P+{\cal M}_{PN},_R\xi^{R}
\bigg)\bigg>_{\cal M}=0
\eea
where $V_{\cal M}^{PN} ={{\delta S_{H}}\over{\delta {\cal M}_{PN}}}$. It is 
understood that ${\cal M}$ has contravariant, covariant and mixed indices. 
Therefore, rules for GCT should be adopted accordingly \cite{gctr}.
 In order to verify
that the ${\cal M}$ derivative of $S_H$ reproduces the vertex operator; one
explicit check is that, for a simple case when we have $G_{MN}$ as the only 
background and set
it to the flat space metric after taking the functional 
derivatives of $S_H$ with respect to $G_{MN}$. Then we reproduce the 
graviton vertex operator. \\
Now we are ready to derive the WI. Note that the infinitesimal parameter,
$\xi ^M(X)$ is arbitrary. Therefore, we may functionally differentiate
(\ref{wi2}) with respect to $\xi^M (X)$ and then set $\xi^M =0$. 
Subsequently, let us take  
functional derivatives of the resulting expression with respect to the  
backgrounds ${\cal M}_{P_iQ_i}({y_i})$, $\{y_i \}$ are the spacetime coordinates,
 and examine the consequences 
\bea
\label{wi3}
&&{\Large \Pi}_{i=1}^{n}{{\delta} \over{{\delta {\cal M}_{P_{i}Q_{i}} }}}
\bigg<\int d^2 \sigma V^{PN}_{\cal M}\bigg[{\cal M}_{PQ}\partial _N\delta
(x-X)
+{\cal M}_{QN}\partial _P\delta(x-X) \nonumber\\
&&+ {\cal M}_{PN},_Q\delta(x-X)\bigg]\bigg>_{\cal M}  =0 
\eea
It is understood that at the end of the operations the backgrounds are
to set the required configurations which is the meaning of $<....>_{\cal M}$
in eq.(\ref{wi1}) - eq.(\ref{wi3}).
These are  desired WI which involve the massless states.
 Let us analyze (\ref{wi3}) more carefully. The 
${\cal M}$-derivatives act in three ways: (i) When it operates on $<...>$ 
action of each derivative brings down a vertex operator 
$\int d^2\sigma V^{P_iQ_i}_{\cal M}\delta(y_i -X(\sigma))$ due to the presence 
of the measure $e^{-iS_H}$ in the definition of $<...>$. Thus we have
eventually $(n+1)$-vertex operators after n-operations. 
(ii) When a derivative acts on the vertex operator, 
$V^{PN}_{\cal M}(X(\sigma))$ it will kill any ${\cal M}$-dependence
in the vertex operator and a corresponding $\delta$-function will appear. 
(iii) The derivatives also act on the $\cal M$-terms 
appearing in (\ref{wi3}) in the square bracket with the $\delta$-functions and
their derivatives. Recall that ${\cal M}$ is an O(d,d) matrix and it   
must be kept in mind while taking functional derivatives.\\
In order to make it more transparent, if we consider the above expression
in the momentum representation, we notice that $(n+1)$-point function 
contracted
with momentum can be expressed in terms of linear combination of 
lower point functions (with contact terms due to the presence of 
$\delta$-functions).
Notice that the WI is $O({\cal D}, {\cal D})$ covariant. Moreover, adopting
the canonical transformation introduced in \cite{mv}, the gauge symmetry
associated with the 2-form field $B_{MN}$ can be revealed, if we choose
${\cal Q}_{\Lambda}=\int d\sigma X'^M\Lambda _M$. Indeed, as has been noted by
Siegel \cite{siegel}, if we define 
\bea
\label{dualgen}
{\cal W}=\pmatrix{\xi ^M \cr \Lambda _M \cr}
\eea
as an $ O({\cal D}, {\cal D})$ vector then we can construct charges (generating
functionals for canonical transformations) as follows: 
\bea
\label{warren}
{\cal Q}_{G}+{\cal Q}_{B}=\int d\sigma {\cal W}^T{\cal V}
\eea
One can check that  operation ${\cal Q}_G +{\cal Q}_B$ on
$S_H$ gives us a relation
\bea
\label{gaugeodd}
(\delta _{Q_G}+\delta _{Q_B})S_H=-\delta _{GCT}S_H-\delta _{Gauge}S_H
\eea
where the second transformation on the {\it r.h.s} is interpreted to be gauge 
variation of background $B_{MN}$ in ${\cal M}$-matrix as
\be
\delta _{Gauge}B_{MN}=\partial _M \Lambda _N - \partial _N \Lambda _M
\ee
$\Lambda _M(X)$ being the vector gauge parameter associated with $B_{MN}$. Thus,
we can derive combined WI, starting from $\Sigma({\cal M})$ and use 
(\ref{dualgen}),  which in its full form will be manifestly duality covariant.
 \\
We now consider compactification of the closed string on d-torii, 
$T^d$, when the
backgrounds along compact directions are independent of those coordinates,
depend only on noncompact coordinates, $X^{\mu}(\sigma\tau),~\mu=1,2..D-1$ and
the compact coordinates are $Y^{\alpha}(\sigma\tau),~ \alpha=1,2,.. d$ 
with $D+d={\cal D}$. 
We adopt the Hassan-Sen \cite{hs,siegel} compactification scheme  where 
the backgrounds, $G_{MN}~ {\rm and} ~ B_{MN}$,
 are decomposed into following block diagonal forms 

\bea
\label{decompose}
G_{MN}(X)=\pmatrix{g_{\mu\nu}(X(\sigma)) & 0\cr 0 & 
G_{\alpha\beta}(X(\sigma)) \cr},~~ 
B_{MN}=\pmatrix{b_{\mu\nu}(X(\sigma)) & 0 \cr 0 & 
B_{\alpha\beta}(X(\sigma)) \cr}
\eea
Thus the action (\ref{action}) can be decomposed into two parts as evident
from (\ref{decompose}). The corresponding canonical Hamiltonian density 
is
\bea
\label{newh}
{\cal H}_c={1\over 2}\bigg({\cal V}^T_1{\bf M}{\cal V}_1+
{\cal V}^T_2{\bf{\tilde M}}{\cal V}_2\bigg)
\eea
which can be written as ${\cal H}_c={\cal H}_1+{\cal H}_2$, the first term
being ${\cal H}_1$ and ${\cal H}_2$ is the second one eq.(\ref{newh}); the
two vectors being 
\bea
\label{oddnew}
{\cal V}_1=\pmatrix{P_{\mu} \cr X'^{\mu}\cr},~~  
{\cal V}_2=\pmatrix{{\tilde P}_{\alpha}\cr Y'^{\alpha}\cr}
\eea
The matrices ${\bf M} ~ {\rm and} ~ {\bf {\tilde M}}$ are defined to be
\bea
\label{newmatrix}
{\bf M}=\pmatrix{g^{\mu\nu} & -g^{\mu\rho}b_{\rho\nu} \cr 
 b_{\mu\rho}g^{\rho\nu} &
g_{\mu\nu}-b_{\mu\rho}g^{\rho\lambda}b_{\lambda\nu} \cr}, ~~
{\bf{\tilde M}}=\pmatrix{G^{\alpha\beta} & -G^{\alpha\gamma}B_{\gamma\beta} \cr
 B_{\alpha\gamma}G^{\gamma\beta} & G_{\alpha\beta}-B_{\alpha\gamma}
G^{\gamma\delta}B_{\delta\beta} \cr}
\eea
Notice that ${\bf M} ~{\rm and}~ {\bf{\tilde M}}$ are 
$O(D,D)~ {\rm and} ~ O(d,d)$ matrices respectively and depend on the noncompact
coordinate $X(\sigma)$. The pair 
${\cal V}_1 ~ {\rm and} ~ {\cal V}_2$ are corresponding two vectors of 
$O(D,D) ~ {\rm and} ~ O(d,d)$. The global $O(D,D)$
transformation is implemented by the $\Omega _1$-matrices and $\Omega _2$
implements the $O(d,d)$ transformation satisfying the properties analogous
to eq.(\ref{omega}); their corresponding metrics
are  
\bea
\label{newmetric}
\eta _1=\pmatrix{0 & 1\cr 1 & 0\cr}, ~~\eta _2=\pmatrix{0 & 1\cr 1 & 0\cr}
\eea
whereas $\eta _1$ is a $2D\times 2D$ matrix, $\eta _2$ is a $2d\times 2d$
matrix. Thus the gravitational WI originating from ${\cal H}_1$  can be
derived following the procedure given above. However, there are massless
scalars(moduli), $G_{\alpha\beta} ~{\rm and} ~ B_{\alpha\beta}$  which appear
in ${\cal H}_2$. These depend on $X^{\mu}$ and therefore, under canonical
transformation (\ref{gravity}) they transform accordingly. The technique
of \cite{maha} can be appropriately used to derive the gravitational WI. 
Similarly, the gauge WI associated with the 2-form field could be obtained
in a straight forward manner. Another important conclusion is that the
canonical Hamiltonian density,${\cal H}_c$, is invariant under global
$O(d,d)$ transformation since 
\bea
\label{odd2}
{\bf{\tilde M}}\rightarrow \Omega _2{\bf{\tilde M}}\Omega _2^T,~~
{\cal V}_2\rightarrow \Omega _2{\cal V}_2
\eea
leaving ${\cal H}_2$ invariant whereas ${\cal H}_1$ is inert under $O(d,d)$
transformation. Furthermore, if we look at the Hamilton's equations of
motion associated with the compact coordinates and their conjugate momenta, 
$\{Y^{\alpha}, P_{\alpha} \}$, we notice
that these are conservation laws (since backgrounds $G_{\alpha\beta} ~{\rm and}
~B_{\alpha\beta}$ depend only on $X^{\mu}$) and the resulting equations 
of motion will
be $O(d,d)$ covariant. Thus we find that the phase space Hamiltonian approach
transparently exposes the duality symmetry.\\
It is worth while to discuss a few more issues relevant to present
investigation. How can we derive the "equations of motion" of the background
fields in this frame work? It can be achieved by resorting to an elegant and
efficient technique proposed in \cite{fmrv}  to obtain the
background equations of motion in Hamiltonian formalism. The quantum 
generators of conformal transformations were constructed by introducing a
generating function  technique. 
For the case at hand, the method of \cite{fmrv} can be suitably 
exploited if we express the ${\cal M}$-matrix in terms of generalized vielbeins
adopted in \cite{ms}: ${\cal M}={\bf V}^T{\bf V}$
\bea
\label{viel}
 {\bf V}=\pmatrix{E^{-1} & -E^{-1}B \cr 0 & E \cr}
\eea
where  $E$, the ${\cal D} \times {\cal D}$ matrix defines the metric
$G_{MN} = E^TE$. We mention in passing that ${\bf V}\in O({\cal D}, {\cal D})$
since ${\bf V}^T \eta {\bf V} =\eta$. Thus the generators ${\cal L}_{\pm}$
will be expressed in terms of $P_M, X'^{M} ~{\rm and}~ {\bf V} $. Now, 
following \cite{fmrv} we can compute anomalies in the quantum algebra of
the generators. These will correspond to known equations of motion as derived
earlier. More importantly, when we consider the case of compactified strings
we note that the constraints obtained in terms of ${\cal H}_1 ~ {\rm and}~
{\cal H}_2$ will give equations of motion for 
$g_{\mu\nu} ~ {\rm and}~ b_{\mu\nu}$ in terms of ${\bf M}$-matrix as well 
as for the 
${\bf{\tilde M}}$-matrix. The  Hamiltonian being $O(d,d)$-invariant the 
equations motion associated with the moduli is expected to be $O(d,d)$ 
covariant since we know that the dimensionally reduced effective action can
expressed in $O(d,d)$ invariant form. \\
We have not discussed the dilaton coupling to the string so far. 
We recall that the dilaton couples to the ghosts and 
their conjugate momenta as was proposed in \cite{mv}
 adopting the arguments of \cite{bns}. Thus the full constraint
algebra can be derived in the Hamiltonian framework
and 
therefore, we can derive the equations of motion for all the massless 
background. The details of such calculations, in the present context, will
presented in a separate publication.\\
Several remarks are in order in what follows. We have argued earlier that 
the phase space measure in the definition of $\Sigma [{\cal M}]$ is
invariant under canonical transformations. However, when the transformed
measure is carefully evaluated in the quantum theory, it might not be invariant
signaling the appearance of an anomaly. We do not have a general prescription
to check the presence of anomalies. In certain cases, the anomaly can be
computed and with specific transformation prescriptions for the backgrounds 
it can be removed \cite{dmp}. However, a general procedure
to derive such anomalies is lacking in this worldshhet approach. \\
The Hamiltonian formalism treats the coordinates and their conjugate momenta
on equal footing in  the $2{\cal D}$-dimensional  the phase space. The duality
symmetry becomes quite transparent in the Hamiltonian descriptions from the
worldsheet point of view. When we considered the Lagrangian formulation,
the equations of motion could be cast in $O(d,d)$ covariant form provided
one introduces dual coordinates for the compact ones and the corresponding
backgrounds are defined suitably in the dual space. 
In the past, it has been suggested that
doubling of the number of coordinates might have underlying deep significance
\cite{duff,wt}. The mathematical formulation of this approach is 
unquestionable; however, the physical significance of such theories are yet
to be fully comprehended. Recently, some progress has been made to compute the
$\beta$-functions in such a worldsheet approach \cite{bt}. Recently interests
in the double field theory formulation have been revived due to a formulation
in the target space \cite{df} where the tensors $G_{MN} ~ {\rm and}~ B_{MN}$
become functions of $2{\cal D}$ variables and the number of indices are also 
doubled. This is a consistent formulation
of the new field theory and it has not found a direct application yet. 
Should one attempt
a Hamiltonian formulation of the 'worldsheet double field theory' \cite{bt},
the phase space will have twice the number of variables contrast to the 
conventional formulations and one will have to suitably define canonical
variables in this frame work. It will be worth while to examine whether such
theories are endowed with any enlarged symmetries.Note, however, that the
$GL(D,R)$ symmetry introduced in \cite{ms} has been found to be important in 
the double field theory  formulation.\\
It has been proposed that excited, massive stated might possess hitherto
undiscovered symmetries \cite{highsym1,sym2,sym3,sym4}. Moreover, some of the 
important properties of dual models, which are inherited by string theory, 
crucially depend on the fact that an infinite tower of states are exchanged
in the scattering processes. Therefore, it is worth while to seek answer to
the question whether the excited massive levels of a string exhibit any
duality-like symmetry. If we examine the issue from the worldsheet view point,
in the $\sigma$-model approach, the (massive) background coupling to the 
string is suppressed by mass term compared to coupling of massless states
on purely dimensional considerations. Therefore, the duality symmetry we 
encounter, in study of the $\sigma$-model action in graviton and 2-form 
potential, will not be unraveled. Similarly, at the level of string effective
action, the dimensionally reduced effective action exhibits duality symmetry
(most commonly known $O(d,d)$) when we assume that the backgrounds do not
depend on compact coordinates and thus ignore the KK modes. Thus the dualities
associated with excited massive modes are to be envisaged from a 
different perspective.
The evolution of string in its excited, massive backgrounds have been studied
in the weak field approximation \cite{sym3,sym4}. One might assume, 
as a simple scenario, that 
the string is  moving in the flat target space and the massive 
backgrounds are
weak. Subsequently construct the vertex operators and demand them to be
conformally invariant which already imposes strong constraints on 
them \cite{highsym1,sym3}. As an illustrative example, 
consider a generic background
coupling of closed bosonic string to its first massive level 
\cite{sym3,highsym1}
\bea
\label{massive1}
F^{(1)}_{MNP}\partial X^{M}\partial X^{N}{\bar\partial}{\bar
\partial}X^{P'},~F^{(2)}_{MN'P'}\partial \partial X^{P}
{\bar\partial}X^{N'}{\bar\partial}X^{P'},~S_{\{MN\}\{P'Q'\}}
\partial X^{M}\partial X^{N}{\bar\partial}X^{P'}{\bar\partial}
X^{Q'} \eea
The backgrounds $F^{(1)}, F^{(2)} ~ {\rm and }~ F^{(3)}$ depend on string
coordinates, $X$, and these term will be suppressed by factor of 
$\alpha '$ compared to the $\sigma$-model action for massless states 
on dimensional 
arguments. The vertex operator for the first excited massive level will
be sum of all such terms (call them vertex functions).
These do not exhaust all possible vertex functions for the first massive
level. The backgrounds fulfill gauge condition and satisfy equations of motion
as a consequence of conformal invariance. If we require them to be $(1,1)$
primary these vertex functions are not independent; it is to be borne in mind
that the stress energy tensors used to compute the weights are taken to be
$T_{++} \sim \partial X\partial X$ and $T_{--}\sim {\bar{\partial}}X
{\bar{\partial }}X$ in the flat target space.\\ 
In order to expose the conjectured duality, we resort to Hamiltonian 
description and assume that the tensors $F^{(i)}$ are spacetime independent
as was first envisaged by Narain, Sarmadi and Witten for the heterotic string
in constant graviton, antisymmetric tensor and gauge field backgrounds. 
Following pertinent points need attentions: (i) A careful reader will notice 
that, when $\{F^{(i)} \}$ are spacetime independent constant tensors, some of
them or their linear combinations might be required to vanish once we demand
that the vertex operator for  the excited massive level 
be $(1,1)$ primary. However, all of them will not
vanish. (ii) When we  study   T-duality symmetry from the worldsheet point
of view in the presence of  massless backgrounds, the resulting equations
of motion are expressed in duality covariant form after incorporating
the dual coordinates \cite{ms}.  It was not essentials for those backgrounds
(i.e. vertex operators) to be $(1,1)$ primary when we are seeking duality
covariant equations of motion. Indeed, conformal invariance lets us decide
which are the admissible background configurations. Therefore, in what
follows, let us analyze how $P\leftrightarrow X'$ duality relates various
(constant) tensor backgrounds. It will be obvious in the sequel that those
tensors which will vanish on imposing $(1,1)$ primary conditions do not
mix with the surviving ones under the duality transformations we are
dealing with. 
Note
that, in flat space $P_M=G^{(0)}_{MN}{\dot X}^N$ where 
$G^{(0)}_{MN}={\rm diag}(+1, -1,-1...)$. In fact we express the vertex 
operators in terms of $P_M, X'^{M}$ for our conveniences here and could
replace $\partial X,~ {\bar {\partial}}X$ by $P\pm X'$ in above expressions
as well.  We would like to consider
following vertex functions which can expressed as linear combination of
appropriate $F$-tensors.
\bea
\label{vertexg}
G^{(1)}_{MNQ}X'^MX'^NX''^Q,~~~G^{(2){MNQ}}P_MP_N{\dot P}_Q,~~~
 G^{(3)Q}_{MN}X'^MX'^N{\dot P}_Q,~~~
 G^{(4)MN}_Q P_MP_NX''^Q,\nonumber \\ G^{(5)}_{MNQR}X'^MX'^NX'^QX'^R,~~~
G^{(6)MNQR}P_MP_NP_QP_R,~~~G_{MN}^{(7)QR}X'^MX'^NP_QP_R
\eea
It is evident that one can construct  more vertex functions for this level;
however, it will suffice to deal with these six for the moment. Note that
$X'{^M} ~{\rm and}~ P_M$ have the same dimensions as it true for the
pair $X''^M ~{\rm and }~{\dot P}_M$.  The simplest form of T-duality the
interchange $\tau \leftrightarrow \sigma$ which implies 
$X'^M \leftrightarrow P_M$
and $X''^M \leftrightarrow{\dot P}_M$. If we desire that the interaction
Hamiltonian consisting of sum of the six terms we have listed above, respect
this duality symmetry, then following relations should hold
\bea
\label{hduality}
G^{(1)}\leftrightarrow G^{(2)},~~G^{(3)}\leftrightarrow G^{(4)},~~
G^{(5)}\leftrightarrow G^{(6)}
\eea
and $G^{(7)}$ gets related to itself with appropriate shuffling of the indices.
This transformation rule generalizes the interchange between $G_{MN}, 
G^{MN} ~{\rm and }~B_{MN}$  for $X'^M\leftrightarrow P_M$ where
$(G+B) \rightarrow (G+B)^{-1}$, alternatively the new metric ${\cal G}$
and the new 2-form ${\cal B}$ (all constants for us) are given by
\bea
\label{newbg}
{\cal G}=(G-BG^{-1}B),~~~{\cal B}=-G^{-1}B((G-BG^{-1}B)
\eea
Notice that these duality relations (\ref{hduality}) hold amongst the
(constant) background tensors of a given level. When we envisage the second
massive excited level of the closed string there will be many more tensors;
however that the vertex operator for the level
 (sum of all such vertex functions) will
be suppressed by a factor ${\alpha '}^2$ 
relative to the first massive level terms.
One might seek answer to the question: Are there larger duality symmetries
associated with massive levels beyond the discrete $P_M\leftrightarrow X'^M$
symmetry considered here? \\
To summarize: we have argued that the WI associated with the
massless excitations of the closed string can be expressed in a duality
covariant manner. It was accomplished by introducing generators of canonical
transformations in the Hamiltonian phase space and defining the generating
functional in path integral formalism.  Furthermore, these generators
\cite{v,mv} can be combined to express in a duality invariant manner. 
The underlying local symmetries are manifest through the Ward identities. 
These WI's
are to be treated as classical expression since anomalies might creep in; 
however,
in certain cases it is possible to compute the anomalies and provide a
prescription to remove them. We outlined a
procedure to compute the quantum constraint algebra in order to derive the
equations of motion for the backgrounds, ${\cal M}$, following the techniques
introduced earlier  in \cite{fmrv}. In fact if one adopts the proposal of
Hohm, Hull and Zwiebach (HHZ) \cite{df} to treat ${\cal M}$ as another $O(d,d)$
spacetime metric (in addition to $\eta$-matrix), then it might facilitate the
computation of $\beta$-functions efficiently. However, it is to be kept in
mind that HHZ's  interpretation was in the context of double field theory.
Therefore, whether truncation to (half) the spacetime variables will be useful
or not is not obvious at this stage. 
We adopted Hassan-Sen
compactification scheme and argued that WI can also be obtained for the 
massless moduli. \\
We have conjectured that there might be duality symmetries associated with
each excited massive level of the closed string. We provided an example how
the constant background tensors should transform to satisfy 
$P\leftrightarrow X'$ interchange. It argued that this type of duality will
persist for higher excited states and the duality relation is to hold for each 
such level. 

\noindent {{\bf Acknowledgments}:} I am grateful to members of 
the String Theory Groups
at Institute of Physics and National Institute of Science Education and
Research (NISER) for fruitful discussions. I am thankful to Yogesh 
Srivastava for carefully
and critically reading the manuscript. I acknowledge very useful
conversations with the Al Mamun at Paris which revived my interests in
worldsheet T-dualities. I would like express my gratitude to
Marios Petropoulos and the Members of CPHT, Ecole Polytechnique for their very 
warm hospitality. This work was primarily supported by
the People of the Republic of India and partly by Indo-French Center for the
Promotion of Advanced Research:
IFCPAR Project No.IFC/4104-2/2010/201.

\end{document}